\documentclass[aps,twocolumn]{revtex4}
\usepackage{graphicx}
\usepackage{epsfig}
\usepackage{epstopdf}
\usepackage{amsfonts}
\usepackage{amssymb}
\usepackage{amsbsy}
\usepackage{amsmath}
\usepackage{mathrsfs}
\usepackage{latexsym}
\usepackage{natbib}
\usepackage{bm}
\usepackage{color}
\usepackage{braket}
\usepackage{slashed}
\usepackage{pgfplots}

\usepackage{tikz}
\usetikzlibrary{shapes,arrows,shadows}

\begin{document}

\title{U(1) symmetry breaking under canonical transformation in real scalar field theory}

\author{Susobhan Mandal}
\email{sm17rs045@iiserkol.ac.in}

\affiliation{ Department of Physical Sciences, 
Indian Institute of Science Education and Research Kolkata,
Mohanpur - 741 246, WB, India }

\date{\today}

\begin{abstract}
In this article, I have considered a real scalar field theory and able to show that under Bogoliubov transformation in infinite volume limit or thermodynamic limit the transformed Hamiltonian no longer invariant under U(1) action defined appropriately as it was before doing transformation. We also have checked this fact by looking at the correlation functions under the action of U(1) group. We suitably defined field operators that are associated with particle production phenomena then we can also show that correlation functions of such field operators also don't follow U(1) invariance, shown in this article. This is a consequence of non-invariance of transformed Hamiltonian under U(1) action. Since, we know Bogoliubov transformation in curved spacetime is equivalent to doing a coordinate transformation, therefore this result directly shows the phenomena of particle production under the affect of gravity since changing coordinate is equivalent to turn on gravity according to Einstein's equivalence principle in GR. I also show that particle production does not take place out of vacuum state but it can happen out of other many-particle states and vacuum state is not an eigenvector of Hamltonian operator in transformed Fock space and vacuum state does not remain vacuum state under time evolution. 
\end{abstract}

\maketitle

\section{Introduction}
We all know field theory describes a system containing infinitely many degrees of freedom and most of the time when we describe a field theoretical system we generally considered a infinite size system or system in thermodynamic limit. This limit is very crucial while we are doing canonical transformation that presereves commutation bracket. One such class is Bogoliubov transformations in scalar field theory which is often used in condensed matter physics \cite{casalbuoni2003lecture}, \cite{timm2012theory}, \cite{chalker2013quantum} and quantum field theory in curved spacetime \citep{birrell1980massive}, \citep{biswas1995particle}, \citep{jacobson2005introduction}. It can be shown and I will show that thermodynamic limit or infinite volume limit make this tranformation impossible and it created two inequivalent representations of two disjoint Fock space which is often used in quantum many body systems. Because of such inequivalent disjoint vector spaces, the operators both in original form and ones after transformation have their own seperate domain to act on states. Therefore, traditional ways of showing particle production \citep{parker2015creation} is not well-defined in thermodynamic limit. However, I actually show in this article that because of under such transformation Hamiltonian is no longer invariant under U(1) action defined in article therefore particle number of the system is not conserved which implies particle production in curved spacetime under gravity.
\vfill\null
\section{Massive scalar free-field theory in hamiltonian description}
\subsection{Massive scalar free-field theory in Minkowski spacetime}
Let's Consider following action
\begin{equation}
\begin{split}
S & =\int\sqrt{-\eta}d^{4}x\Big[\frac{1}{2}(\partial\phi(x))^{2}-\frac{1}{2}m^{2}\phi^{2}(x)\Big]\\
\eta & =\text{diag}(1,-1,-1,-1)
\end{split}
\end{equation} 
We can write field operators in following way
\begin{equation}
\hat{\phi}(x) =\int\frac{d^{3}k}{\sqrt{(2\pi)^{3}2\omega_{\vec{k}}}}[\hat{a}_{\vec{k}}e^{-ik.x}+\hat{a}_{\vec{k}}^{\dagger}e^{ik.x}]
\end{equation} 
Here canonical conjugate momentum field operator is $\hat{\pi}=\dot{\hat{\phi}}$. And one can check from canonical commutation relation
\begin{equation}
[\hat{\phi}(x),\hat{\pi}(y)]_{x^{0}=y^{0}} =i\delta^{(3)}(\vec{x}-\vec{y})
\end{equation} 
following algebra between creation-annihilation operators
\begin{equation}
\begin{split}
[\hat{a}_{\vec{k}},\hat{a}_{\vec{k}'}] & =0=[\hat{a}_{\vec{k}}^{\dagger},\hat{a}_{\vec{k}'}^{\dagger}]\\
[\hat{a}_{\vec{k}},\hat{a}_{\vec{k}'}^{\dagger}] & =\delta^{(3)}(\vec{k}-\vec{k}')
\end{split}
\end{equation} 
And Hamiltonian operator for this system is following \cite{book:14817}
\begin{equation}
\begin{split}
\hat{H} & =\int d^{3}x\Big[\frac{1}{2}\hat{\pi}^{2}(x)+\frac{1}{2}(\vec{\nabla}\phi(x))^{2}+\frac{1}{2}m^{2}\phi^{2}(x)\Big]\\
 & =\int d^{3}k\frac{1}{2}\omega_{\vec{k}}[\hat{a}_{\vec{k}}\hat{a}_{\vec{k}}^{\dagger}+\hat{a}_{\vec{k}}^{\dagger}\hat{a}_{\vec{k}}]\\
 & \equiv\int d^{3}k\omega_{\vec{k}}\hat{a}_{\vec{k}}^{\dagger}\hat{a}_{\vec{k}}, \ \omega_{\vec{k}}=\sqrt{\vec{k}^{2}+m^{2}}
\end{split}
\end{equation} 
The above expression can also be written using the definition of number operators
\begin{equation}
\begin{split}
\hat{N}_{\vec{k}} & =\hat{a}_{\vec{k}}^{\dagger}\hat{a}_{\vec{k}}\\
\implies\hat{H} & =\int d^{3}k \ \omega_{\vec{k}}\hat{N}_{\vec{k}}
\end{split}
\end{equation} 
Note that Hamiltonian of this theory is invariant both global $U(1)$-group action which is following
\begin{equation}
\begin{split}
\hat{a}_{\vec{k}} & \rightarrow e^{i\Theta}\hat{a}_{\vec{k}}\\
\hat{a}_{\vec{k}}^{\dagger} & \rightarrow e^{-i\Theta}\hat{a}_{\vec{k}}^{\dagger}
\end{split}
\end{equation} 
and local $U(1)$-group action which is following
\begin{equation}
\begin{split}
\hat{a}_{\vec{k}} & \rightarrow e^{i\Theta_{\vec{k}}}\hat{a}_{\vec{k}}\\
\hat{a}_{\vec{k}}^{\dagger} & \rightarrow e^{-i\Theta_{\vec{k}}}\hat{a}_{\vec{k}}^{\dagger}
\end{split}
\end{equation} 
And vacuum of this theory is state $\ket{0}$ which is such that
\begin{equation}
\hat{a}_{\vec{k}}\ket{0}  =0, \ \forall\vec{k}
\end{equation} 
And all the single and multi-particle states can be constructed using operation of creation operators on the vacuum state. 
\subsection{Massive Scalar Free-Field Theory in different frame}
Now we want to move to a different frame which is non-inertial and w.r.t an observer from this frame action can be written down using minimal prescription
\begin{equation}
\begin{split}
S & =\int\sqrt{-g}d^{4}x\Big[\frac{1}{2}(\partial\phi(x))^{2}-\frac{1}{2}m^{2}\phi^{2}(x)\Big]\\
(\partial\phi(x))^{2} & =g^{\mu\nu}(x)\partial_{\mu}\phi(x)\partial_{\nu}\phi(x)
\end{split}
\end{equation} 
where the metric $g_{\mu\nu}(x)$ is non-trivial.\\[5pt]
Here we choose the mode functions to be solutions of Euler-Lagrange equation which is of following form
\begin{equation}
\frac{1}{\sqrt{-g}}\partial_{\mu}(\sqrt{-g}g^{\mu\nu}\partial_{\nu}\phi)+m^{2}\phi=0
\end{equation}
Let $\{f_{n}(x)\}$ be complete set of such mode functions which solve above equation.\\[5pt]
We define an inner product between 2 such mode function in following way
\begin{equation}
(f,g)\equiv i\int\sqrt{-g}d^{3}x \ f^{*}(x)[g^{0\nu}\overrightarrow{\partial}_{\nu}-g^{0\nu}\overleftarrow{\partial}_{\nu}]g(x)
\end{equation} 
And one can show that this inner product is time-translation invariant \cite{book:274311}
\begin{equation}
\begin{split}
\partial_{0}(f,g) & = i\int d^{3}x \ \partial_{0}\Big[\sqrt{-g}f^{*}(x)[g^{0\nu}\overrightarrow{\partial}_{\nu}-g^{0\nu}\overleftarrow{\partial}_{\nu}]g(x)\Big]\\
 & =i\int d^{3}x \ \partial_{\mu}\Big[\sqrt{-g}f^{*}(x)[g^{\mu\nu}\overrightarrow{\partial}_{\nu}-g^{\mu\nu}\overleftarrow{\partial}_{\nu}]g(x)\Big]\\
 & -i\int d^{3}x \ \partial_{i}\Big[\sqrt{-g}f^{*}(x)[g^{i\nu}\overrightarrow{\partial}_{\nu}-g^{i\nu}\overleftarrow{\partial}_{\nu}]g(x)\Big]\\
=i\int d^{3}x \ & \Big[f^{*}(x)\partial_{\mu}[\sqrt{-g}g^{\mu\nu}\overrightarrow{\partial}_{\nu}]-\partial_{\mu}[\sqrt{-g}g^{\mu\nu}f^{*}(x)\overleftarrow{\partial}_{\nu}]g(x)\Big]\\
 & -i\int d^{3}x \ \partial_{i}\Big[\sqrt{-g}f^{*}(x)[g^{i\nu}\overrightarrow{\partial}_{\nu}-g^{i\nu}\overleftarrow{\partial}_{\nu}]g(x)\Big]\\
 =0
\end{split}
\end{equation} 
where we throw away the surface term because we have assumed that fields vanish at surface near spatial infinity and consider Euler-Lagrange equation followed by the modes.\\[5pt] 
Similar definition is also there in first frame with only difference is the metric is Minkowski. So, we label the inner product in first and second frame to be 1 and 2 respectively.\\[5pt]
In this frame also we can write the field operators in following way
\begin{equation}
\hat{\phi}'(y) = \sum_{n}[\hat{b}_{n}f_{n}(y)+\hat{b}_{n}^{\dagger}f_{n}^{*}(y)]
\end{equation}
Now consider a point in spacetime which has different coordinates w.r.t two frames and let call them $x$ and $x'$ respectively in initial and final frame. Since, we are dealing with scalar field theory which has a nice property which is its tranformation property under general coordinate transformation
\begin{equation}
\hat{\phi}'(x') =\hat{\phi}(x)
\end{equation}
Using the definition of inner-product one can define a Bogoliubov type transformation which is defined in next subsection. And for convenience we choose momentum mode decomposition of fields from next section onwards by consiering spacetime which has spatial translational invariant for example spatially flat FRW spacetime \citep{book:274311}.
\subsection{Bogoliubov transformation}
To show the Bogoliubov transformation \cite{jacobson2005introduction}, \cite{sato1994bogoliubov}, we consider a real scalar field with modes $\{\hat{a}_{\vec{k}}\}$. And they satisfy following algebra
\begin{equation}
[\hat{a}_{\vec{k}},\hat{a}_{\vec{q}}^{\dagger}]=\delta^{(3)}(\vec{k}-\vec{q})
\end{equation}
and rest of the commutator brackets are zero.\\[5pt]
With these one can construct Fock space $\mathcal{H}[a]$ with repeated applications of $\hat{a}_{\vec{k}}^{\dagger}$s on vacuum state denoted by $\ket{0}$ defined by
\begin{equation}
\hat{a}_{\vec{k}}\ket{0}=0, \ \forall\vec{k}
\end{equation}
Let us now consider the following Bogoliubov transformation(this is not most general transformation because most general transformation also can mix different momentum modes)
\begin{equation}
\begin{split}
\hat{c}_{\vec{k}}(\theta) & =\cosh\theta_{\vec{k}}\hat{a}_{\vec{k}}-\sinh\theta_{\vec{k}}\hat{a}_{-\vec{k}}^{\dagger}
\end{split}
\end{equation}
With these transformations in hand, one can check that new operators also satisfy
\begin{equation}
[\hat{c}_{\vec{k}}(\theta),\hat{c}_{\vec{q}}^{\dagger}(\theta)]=\delta^{(3)}(\vec{k}-\vec{q})
\end{equation}
and all other commutators vanishing.\\[5pt]
Now we consider vacuum relative to the operators $\{\hat{c}_{\vec{k}}(\theta)$, denoted by $\ket{0(\theta)}$ and defined by
\begin{equation}
\hat{c}_{\vec{k}}(\theta)\ket{0(\theta)}=0, \ \forall\vec{k}
\end{equation}
and we construct new Fock space representation $\mathcal{H}(c)$ by repeated application of $\{\hat{c}_{\vec{k}}^{\dagger}(\theta)$ on the vacuum state $\ket{0(\theta)}$.\\[5pt]
If now we assume the existence of an unitary operator $G(\theta)$ which generates the transformation
\begin{equation}
U(\theta)\hat{a}_{\vec{k}}U^{-1}(\theta)=\hat{c}_{\vec{k}}(\theta)
\end{equation}
where $U(\theta)=e^{iG(\theta)}$, then one can explicitly check that
\begin{equation}
G(\theta)=i\int d^{3}k \ \theta_{\vec{k}}(\hat{a}_{\vec{k}}\hat{a}_{-\vec{k}}-\hat{a}_{-\vec{k}}^{\dagger}\hat{a}_{\vec{k}}^{\dagger})
\end{equation}
And therefore, one can write the transformation operator in following way
\begin{equation}
\begin{split}
U(\theta) & =e^{-\delta^{(3)}(0)\int d^{3}l\ln\cosh\theta_{\vec{k}}}e^{\int d^{3}k\tanh\theta_{\vec{k}}\hat{a}_{\vec{k}}^{\dagger}\hat{a}_{-\vec{k}}^{\dagger}}\\
 & \times^{-\int d^{3}k\tanh\theta_{\vec{k}}\hat{a}_{-\vec{k}}\hat{a}_{\vec{k}}}
\end{split}
\end{equation}
then we have 
\begin{equation}
\ket{0(\theta)}=e^{-\delta^{(3)}(0)\int d^{3}k\ln\cosh\theta_{\vec{k}}}e^{\int d^{3}k\tanh\theta_{\vec{k}}\hat{a}_{\vec{k}}^{\dagger}\hat{a}_{-\vec{k}}^{\dagger}}\ket{0}
\end{equation}
where $\delta^{(3)}(0)$ in discrete limit is volume V. Therefore, unless $V<\infty$ and $\int d^{3}k\ln\cosh\theta_{\vec{k}}<\infty$, state $\ket{0(\theta)}$ does not belong to $\mathcal{H}(a)$ Fock space therefore, we can say that these two representations are inequivalent \cite{stepanian2013unitary}.\\[5pt]
Note that transformation between operators are well-defined for any volume of the system, so once we defined the transformation we take the limit $V\rightarrow\infty$, for which transformation between operators are still well-defined but new and old Fock space becomes disjoint. Therefore, new and old creation and annihilation operators have seperate Hilbert space to act one or in otherwords theor domains become disjoint. This construction will be assumed from next section onwards. 
\subsection{Hamiltonian under Bogoliubov transformation}
Now let's look at the Hamiltonian under Bogoliubov transformation but before that we need the inverse Bogoliubov transformation
\begin{equation}
\begin{split}
\hat{c}_{\vec{k}}(\theta) & =\cosh\theta_{\vec{k}}\hat{a}_{\vec{k}}-\sinh\theta_{\vec{k}}\hat{a}_{-\vec{k}}^{\dagger}\\
\hat{c}_{\vec{k}}^{\dagger}(\theta) & =\cosh\theta_{\vec{k}}\hat{a}_{\vec{k}}^{\dagger}-\sinh\theta_{\vec{k}}\hat{a}_{-\vec{k}}\\
\implies\hat{a}_{\vec{k}} & =\cosh\theta_{\vec{k}}\hat{c}_{\vec{k}}+\sinh\theta_{\vec{k}}\hat{c}_{-\vec{k}}^{\dagger}\\
\hat{a}_{\vec{k}}^{\dagger} & =\cosh\theta_{\vec{k}}\hat{c}_{\vec{k}}^{\dagger}+\sinh\theta_{\vec{k}}\hat{c}_{-\vec{k}}
\end{split}
\end{equation}
Using this we can write
\begin{equation}
\begin{split}
\hat{H} & =\int d^{3}k \ \varepsilon_{\vec{k}}\hat{a}_{\vec{k}}^{\dagger}\hat{a}_{\vec{k}}\\
 & =\int d^{3}k \ \varepsilon_{\vec{k}}[\cosh\theta_{\vec{k}}\hat{c}_{\vec{k}}^{\dagger}+\sinh\theta_{\vec{k}}\hat{c}_{-\vec{k}}]\\
 & \times[\cosh\theta_{\vec{k}}\hat{c}_{\vec{k}}+\sinh\theta_{\vec{k}}\hat{c}_{-\vec{k}}^{\dagger}]\\
 & =\int d^{3}k \ \varepsilon_{\vec{k}}[\cosh^{2}\theta_{\vec{k}}\hat{c}_{\vec{k}}^{\dagger}\hat{c}_{\vec{k}}+\sinh^{2}\theta_{\vec{k}}\hat{c}_{-\vec{k}}\hat{c}_{-\vec{k}}^{\dagger}\\
 & +\sinh\theta_{\vec{k}}\cosh\theta_{\vec{k}}(\hat{c}_{-\vec{k}}\hat{c}_{\vec{k}}+\hat{c}_{\vec{k}}^{\dagger}\hat{c}_{-\vec{k}}^{\dagger})]\\
 & \simeq\int d^{3}k \ \varepsilon_{\vec{k}}[(\cosh^{2}\theta_{\vec{k}}+\sinh^{2}\theta_{-\vec{k}})\hat{c}_{\vec{k}}^{\dagger}\hat{c}_{\vec{k}}\\
 & +\sinh\theta_{\vec{k}}\cosh\theta_{\vec{k}}(\hat{c}_{-\vec{k}}\hat{c}_{\vec{k}}+\hat{c}_{\vec{k}}^{\dagger}\hat{c}_{-\vec{k}}^{\dagger})]
\end{split}
\end{equation}
Note that under Bogoliubov transformation Hamiltonian breaks $U(1)$ invariance in new Fock space representation. Also note that this Hamiltonian looks similar to BCS superconductor Hamiltonian in mean-field approximation \cite{timm2012theory}, \cite{casalbuoni2003lecture} but here we have non-trivial coefficients attached to the operators.\\[5pt]
Bow we are willing to write the Hamiltonian in matrix form and to do that we choose a representation w.r.t basis of the form $\begin{pmatrix}
\hat{c}_{\vec{k}}\\
\hat{c}_{-\vec{k}}^{\dagger}
\end{pmatrix}, \ \forall\vec{k}$. Using this representation we can write the Hamiltonian in following form
\begin{equation}
\begin{split}
\hat{H}=\int d^{3}k\varepsilon_{\vec{k}} & \begin{pmatrix}
\hat{c}_{\vec{k}}^{\dagger} & \hat{c}_{-\vec{k}}
\end{pmatrix}\begin{bmatrix}
\cosh^{2}\theta_{\vec{k}} & \sinh\theta_{\vec{k}}\cosh\theta_{\vec{k}}\\
\sinh\theta_{\vec{k}}\cosh\theta_{\vec{k}} & \sinh^{2}\theta_{\vec{k}}
\end{bmatrix}\\
 & \times\begin{pmatrix}
\hat{c}_{\vec{k}}\\
\hat{c}_{-\vec{k}}^{\dagger}
\end{pmatrix}
\end{split}
\end{equation}
This representation will help us latter in finding off-diagonal components of 2-point correlation function or Green's function.\\[5pt]
Before we proceed to next subsection where we start discussion of coherent states and partition function we want to make some comments
\begin{itemize}
\item First of all, note that new vacuum state that we got in new Fock space representation made out of combination of states of pair of $a-$ particles with opposite momentum, which can be seen directly from the mathematical definition of new vacuum state in terms of old vacuum state.
\item Secondly, since Hamiltonian in new Fock space representation breaks the $U(1)$-invariance therefore we expect particle number is not conserved which will also reflect in correlation functions which we will show later. 
\end{itemize}
\section{Correlation functions}
\subsection{Description of Coherent states}
Let's strat with simple harmonic oscillator description first where we have as usual following algebra
\begin{equation}
[\hat{a},\hat{a}^{\dagger}]=1, \ [\hat{a},\hat{a}]=0=[\hat{a}^{\dagger},\hat{a}^{\dagger}]
\end{equation}
We define coherent states to be the states in Hilbert space which are eigenstates of annihilation operator $\hat{a}$ and defined in following way
\begin{equation}
\ket{z}=e^{z\hat{a}^{\dagger}}\ket{0}
\end{equation}
where $z$ is a complex number and $\ket{0}$ is vacuum state or ground state in this case. One can easily check that
\begin{equation}
\hat{a}\ket{z}=\hat{a}e^{z\hat{a}^{\dagger}}\ket{0}=z\ket{z}
\end{equation}
and similarly correspong dual state can be written as
\begin{equation}
\bra{z}=\bra{0}e^{\bar{z}\hat{a}}
\end{equation}
Similarly with little bit algebra one can show that
\begin{equation}
\braket{z|z'}=e^{\bar{z}z'}
\end{equation}
One can similarly show the resolution of identity
\begin{equation}
\hat{I}=\int\frac{dz d\bar{z}}{2\pi i}e^{-z\bar{z}}\ket{z}\bra{z}
\end{equation}
And for any normal ordered operator $A(\hat{a}^{\dagger},\hat{a})$ one can show easily that
\begin{equation}
\bra{z}A(\hat{a}^{\dagger},\hat{a})\ket{z'}=A(\bar{z},z')e^{\bar{z}z'}
\end{equation}
\subsubsection{Number operator as generator of phase}
In above construction number operator is defined as
\begin{equation}
\hat{N}=\hat{a}^{\dagger}\hat{a}
\end{equation}
Now consider a coherent state $\ket{z}$ such that $\hat{a}\ket{z}=z\ket{z}$ where $z$ is some complex number. Then we look at the state $e^{-i\hat{N}\theta}\ket{z}$
\begin{equation}
\begin{split}
\hat{a}e^{-i\hat{N}\theta}\ket{z} & =e^{-i\hat{N}\theta}e^{i\hat{N}\theta}\hat{a}e^{-i\hat{N}\theta}\ket{z}\\
=e^{-i\hat{N}\theta} & \Big[\hat{a}+i\theta[\hat{N},\hat{a}]+\frac{(i\theta)^{2}}{2}[\hat{N},[\hat{N},\hat{a}]]+\ldots\Big]\ket{z}\\
 =e^{-i\hat{N}\theta} & e^{-i\theta}\hat{a}\ket{z}=z e^{-i\theta}e^{-i\hat{N}\theta}\ket{z}\\
\implies e^{-i\hat{N}\theta}\ket{z} & \propto\ket{z e^{-i\theta}}
\end{split}
\end{equation}
Therefore, we can see that number operator $\hat{N}$ is the generator of complex phase for coherent states. We will come to this point again later in this article. 
\subsection{Path integral using coherent states}
We will want to compute the matrix elements of the evolution operator $\hat{U}$ defined by
\begin{equation}
\hat{U}(t_{f},t_{i})=e^{-\frac{i}{\hbar}T\hat{H}(\hat{a}^{\dagger},\hat{a})}
\end{equation}
where $T=(t_{f}-t_{i})$ and $\hat{H}(\hat{a}^{\dagger},\hat{a})$ is the Hamiltonian operator of the system in normal ordered form. Thus, if $\ket{i}$ and $\ket{f}$ denote two arbitrary initial and final states, we can write the matrix element of $\hat{U}(t_{f},t_{i})$ as $\bra{f}\hat{U}(t_{f},t_{i})\ket{i}$. Now we split the whole time interval into $N$-equal segments with $N\rightarrow\infty$, then each segment has length of $\epsilon\rightarrow0^{+}$, then using the first order approximation we can write
\begin{equation}
\bra{f}\hat{U}(t_{f},t_{i})\ket{i}=\lim_{\epsilon\rightarrow0^{+}}\lim_{N\rightarrow\infty}\bra{f}\left(1-\frac{i}{\hbar}\epsilon\hat{H}(\hat{a}^{\dagger},\hat{a})\right)^{N}\ket{i}
\end{equation}
Then using an overcomplete set $\{\ket{z_{j}}\}$ at each time $t_{j}$ where $j=1,\ldots,N$ and using the resolution of identity for coherent states one can show that
\begin{equation}
\begin{split}
\lim_{\epsilon\rightarrow0^{+}}\lim_{N\rightarrow\infty}\bra{f} & \left(1-\frac{i}{\hbar}\epsilon\hat{H}(\hat{a}^{\dagger},\hat{a})\right)^{N}\ket{i}\\
=\int & \left(\prod_{j=1}^{N}\frac{dz_{j}d\bar{z}_{j}}{2\pi i}\right)e^{-\sum_{j}|z_{j}|^{2}}\\
\times\Big[\prod_{k=1}^{N-1}\bra{z_{k+1}} & \left(1-\frac{i}{\hbar}\epsilon\hat{H}(\hat{a}^{\dagger},\hat{a})\right)\ket{z_{k}}\Big]\\
\times\bra{f}\left(1-\frac{i}{\hbar}\epsilon\hat{H}(\hat{a}^{\dagger},\hat{a})\right)\ket{z_{N}} & \bra{z_{1}}\left(1-\frac{i}{\hbar}\epsilon\hat{H}(\hat{a}^{\dagger},\hat{a})\right)\ket{i}
\end{split}
\end{equation}
In the limit $\epsilon\rightarrow0^{+}$ these matrix elements become
\begin{equation}
\begin{split}
\bra{z_{k+1}}\left(1-\frac{i}{\hbar}\epsilon\hat{H}(\hat{a}^{\dagger},\hat{a})\right)\ket{z_{k}} & =\braket{z_{k+1}|z_{k}}\\
 & \times\Big[1-\frac{i}{\hbar}\epsilon\hat{H}(\bar{z}_{k+1},z_{k})\Big]
\end{split}
\end{equation}
And using the above information we can write
\begin{equation}
\begin{split}
\bra{f}\hat{U}(t_{f},t_{i})\ket{i} & =\int\left(\prod_{j=1}^{N}\frac{dz_{j}d\bar{z}_{j}}{2\pi i}\right)e^{-\sum_{j}|z_{j}|^{2}}e^{\sum_{j=1}^{N-1}\bar{z}_{j+1}z_{j}}\\
\times & \prod_{k=1}^{N-1}\Big[1-\frac{i}{\hbar}\epsilon\hat{H}(\bar{z}_{k+1},z_{k})\Big]\braket{f|z_{N}}\braket{z_{1}|i}\\
 & \times\Big[1-\frac{i\epsilon}{\hbar}\frac{\bra{f}\hat{H}\ket{z_{N}}}{\braket{f|z_{N}}}\Big]\Big[1-\frac{i\epsilon}{\hbar}\frac{\bra{z_{1}}\hat{H}\ket{i}}{\braket{z_{1}|i}}\Big]\\
=\int\mathcal{D}z\mathcal{D}\bar{z} & e^{\frac{i}{\hbar}\int_{t_{i}}^{t_{f}}dt\Big[\frac{\hbar}{2i}(z\partial_{t}\bar{z}-\bar{z}\partial_{t}z)-H(\bar{z},z)\Big]}\\
\times & e^{\frac{1}{2}(|z_{i}|^{2}+|z_{f}|^{2})}\bar{\psi}_{f}(z_{f})\psi_{i}(\bar{z}_{i})
\end{split}
\end{equation}
where we have used the fact that
\begin{equation}
\begin{split}
\bra{f} & =\int\frac{dz_{f}d\bar{z}_{f}}{2\pi i}e^{-|z_{f}|^{2}}\bar{\psi}_{f}(z_{f})\bra{z_{f}}\\
\ket{i} & =\int\frac{dz_{i}d\bar{z}_{i}}{2\pi i}e^{-|z_{i}|^{2}}\psi_{i}(\bar{z}_{i})\ket{z_{i}}
\end{split}
\end{equation}
\subsection{Extend the path integral to field theory}
As we have seen in free-field theory we have harmonic oscillators with different momentum modes for which we have Hamiltonian in old Fock space as
\begin{equation}
\hat{H}=\int d^{3}k \ \varepsilon_{\vec{k}}\hat{c}_{\vec{k}}^{\dagger}\hat{
c}_{\vec{k}}
\end{equation}
Now we define states \cite{book:18638}
\begin{equation}
\ket{\phi}=e^{\int d^{3}k\phi(\vec{k})\hat{c}_{\vec{k}}^{\dagger}}\ket{0}
\end{equation}
and one can now check that
\begin{equation}
\hat{c}_{\vec{l}}\ket{\phi}=\phi({\vec{l}})\ket{\phi}
\end{equation}
as well as obeying the resolution of the identity in this space of states
\begin{equation}
\hat{I}=\int\mathcal{D}\phi\mathcal{D}\bar{\phi}e^{-\int d^{3}k|\phi(\vec{k})|^{2}}\ket{\phi}\bra{\phi}
\end{equation}
Therefore, using above extended quantities one can easily show that
\begin{equation}
\begin{split}
\bra{f} & e^{-\frac{i}{\hbar}T\hat{H}}\ket{i}\\
=\int\mathcal{D}\phi\mathcal{D}\bar{\phi} & \bar{\Psi}_{f}(\phi(t_{f},\vec{k}))\Psi(\bar{\phi}(t_{i},\vec{k}))e^{\frac{1}{2}\int d^{3}k(|\phi(t_{i},\vec{k})|^{2}+|\phi(t_{f},\vec{k})|^{2})}\\
\times & e^{\frac{i}{\hbar}\int_{t_{i}}^{t_{f}}\int d^{3}k\Big[\frac{\hbar}{2i}(\phi(t,\vec{k})\partial_{t}\bar{\phi}(t,\vec{k})-\bar{\phi}(t,\vec{k})\partial_{t}\phi(t,\vec{k}))-\varepsilon_{\vec{k}}\bar{\phi}(t,\vec{k})\phi(t,\vec{k})\Big]}\\
=\int\mathcal{D}\phi\mathcal{D}\bar{\phi} & \ e^{\frac{i}{\hbar}\int_{t_{i}}^{t_{f}}\int d^{3}k\Big[i\hbar\bar{\phi}(t,\vec{k})\partial_{t}\phi(t,\vec{k})-\varepsilon_{\vec{k}}\bar{\phi}(t,\vec{k})\phi(t,\vec{k})\Big]}\\
\times & \bar{\Psi}_{f}(\phi(t_{f},\vec{k}))\Psi(\bar{\phi}(t_{i},\vec{k}))e^{\frac{1}{2}\int d^{3}k(|\phi(t_{i},\vec{k})|^{2}+|\phi(t_{f},\vec{k})|^{2})}
\end{split}
\end{equation}
Now we consider partition function in grand canonical ensemble, defined by
\begin{equation}
\mathcal{Z}=\text{Tr}e^{-\beta(\hat{H}-\mu\hat{N})}
\end{equation}
which we want to evaluate using the path integral formalism where we extend the formalism in following way
\begin{itemize}
\item $\ket{i}=\ket{f}$ and arbitrary.
\item summing over boundary states
\item wick rotation to imaginary time $t\rightarrow-i\tau$ with time span $T\rightarrow-i\beta\hbar$ \cite{book:428978}.
\end{itemize}
The result will be following \cite{laine2016basics}, \cite{yang2011introduction}
\begin{equation}
\mathcal{Z}=\int\mathcal{D}\phi\mathcal{D}\bar{\phi} \ e^{-S_{E}(\phi,\bar{\phi})}
\end{equation}
where 
\begin{equation}
S_{E}(\phi,\bar{\phi})=\frac{1}{\hbar}\int_{0}^{\beta\hbar}d\tau\int d^{3}k\bar{\phi}(\tau,\vec{k})\Big[\hbar\partial_{\tau}-\xi_{\vec{k}}\Big]\phi(\tau,\vec{k})
\end{equation}
with $\xi_{\vec{k}}=\varepsilon_{\vec{k}}-\mu$ and with periodic boundary condition $\phi(\tau,\vec{k})=\phi(\tau+\beta\hbar,\vec{k})$. This requiremnet suggests that we can decompose $\phi(\tau,\vec{k})$ in following way
\begin{equation}
\phi(\tau,\vec{k})=\sum_{n}e^{i\omega_{n}\tau}\phi_{n}(\vec{k})
\end{equation}
where $\omega_{n}=\frac{2\pi n}{\beta\hbar}$, which are known as matsubara frequencies.
\subsection{Correlation function in curved spacetime under Bogoliubov transformation}
Recall that after doing Bogoliubov transformation new Hamiltonian for real scalar free-fiel theory was of the form
\begin{equation}
\begin{split}
\hat{H} & =\int d^{3}k \ \varepsilon_{\vec{k}}[\cosh^{2}\theta_{\vec{k}}\hat{c}_{\vec{k}}^{\dagger}\hat{c}_{\vec{k}}+\sinh^{2}\theta_{\vec{k}}\hat{c}_{-\vec{k}}\hat{c}_{-\vec{k}}^{\dagger}\\
 & +\sinh\theta_{\vec{k}}\cosh\theta_{\vec{k}}(\hat{c}_{-\vec{k}}\hat{c}_{\vec{k}}+\hat{c}_{\vec{k}}^{\dagger}\hat{c}_{-\vec{k}}^{\dagger})]\\
 & \simeq\int d^{3}k \ \varepsilon_{\vec{k}}[(\cosh^{2}\theta_{\vec{k}}+\sinh^{2}\theta_{-\vec{k}})\hat{c}_{\vec{k}}^{\dagger}\hat{c}_{\vec{k}}\\
 & +\sinh\theta_{\vec{k}}\cosh\theta_{\vec{k}}(\hat{c}_{-\vec{k}}\hat{c}_{\vec{k}}+\hat{c}_{\vec{k}}^{\dagger}\hat{c}_{-\vec{k}}^{\dagger})]
\end{split}
\end{equation} 
Therefore, for this Hamiltonian, the Euclidean action in the exponent of path integral description of partition function will be
\begin{equation}
\begin{split}
S_{E}(\bar{\phi},\phi) & =\frac{1}{\hbar}\int_{0}^{\beta\hbar}d\tau\int d^{3}k\\
 & \Big[\bar{\phi}(\tau,\vec{k})(\hbar\partial_{\tau}-(\cosh^{2}\theta_{\vec{k}}+\sinh^{2}\theta_{\vec{k}})\xi_{\vec{k}})\phi(\tau,-\vec{k})\\
-\xi_{\vec{k}}\sinh\theta_{\vec{k}} & \cosh\theta_{\vec{k}}(\phi(\tau,\vec{k})\phi(\tau,-\vec{k})+\bar{\phi}(\tau,\vec{k})\bar{\phi}(\tau,-\vec{k}))\Big]
\end{split}
\end{equation}
which can be written in matrix representation in following way(denoting $\chi_{\vec{k}}=\cosh^{2}\theta_{\vec{k}}+\sinh^{2}\theta_{\vec{k}}, \ \eta_{\vec{k}}=\cosh\theta_{\vec{k}}\sinh\theta_{\vec{k}}$ and we assumed $\chi_{\vec{k}}=\chi_{-\vec{k}}, \ \eta_{\vec{k}}=\eta_{-\vec{k}}$)
\begin{equation}
\begin{split}
S_{E}(\bar{\phi},\phi)=\frac{1}{\hbar}\int_{0}^{\beta\hbar}d\tau\int d^{3}k & \begin{pmatrix}
\bar{\phi}(\tau,\vec{k}) & \phi(\tau,-\vec{k}) 
\end{pmatrix}\\
\times\begin{bmatrix}
\frac{1}{2}(\hbar\partial_{\tau}-\xi_{\vec{k}}\chi_{\vec{k}}) & \xi_{\vec{k}}\eta_{\vec{k}}\\
\xi_{\vec{k}}\eta_{\vec{k}} & \frac{1}{2}(\hbar\overleftarrow{\partial}_{\tau}-\xi_{\vec{k}}\chi_{\vec{k}})
\end{bmatrix}
 & \times\begin{pmatrix}
\phi(\tau,\vec{k})\\
\bar{\phi}(\tau,-\vec{k})
\end{pmatrix}\\
=\beta\sum_{n}\int d^{3}k \begin{pmatrix}
\bar{\phi}(\omega_{n},\vec{k}) & \phi(-\omega_{n},-\vec{k}) 
\end{pmatrix} & \times\\
\begin{bmatrix}
\frac{1}{2}(i\hbar\omega_{n}-\xi_{\vec{k}}\chi_{\vec{k}}) & \xi_{\vec{k}}\eta_{\vec{k}}\\
\xi_{\vec{k}}\eta_{\vec{k}} & \frac{1}{2}(-i\hbar\omega_{n}-\xi_{\vec{k}}\chi_{\vec{k}})
\end{bmatrix} & \times\begin{pmatrix}
\phi(\omega_{n},\vec{k})\\
\bar{\phi}(-\omega_{n},-\vec{k})
\end{pmatrix} 
\end{split}
\end{equation} 
Now we define generating functional to get the correlation functions of any order. It is defined as(from now on we consider $\hbar=1$)
\begin{equation}
\begin{split}
\mathcal{Z}[J,\bar{J}] & =\frac{1}{\mathcal{Z}}\int\mathcal{D}\phi \ \mathcal{D}\bar{\phi} \ e^{-S_{E}(\phi,\bar{\phi})}\\
 & \times e^{\sum_{n}\int d^{3}k(\bar{J}(\omega_{n},\vec{k})\phi(\omega_{n},\vec{k})+J(\omega_{n},\vec{k})\bar{\phi}(\omega_{n},\vec{k}))}\\
 & =e^{\bar{J}\mathcal{G}J}
\end{split}
\end{equation}
where $\bar{J}\mathcal{G}J$ denotes a matrix multiplication with sum over modes and $\mathcal{G}$ is the propagator matrix or 2-point function matrix. Let's write down $\bar{J}\mathcal{G}J$ explicitly
\begin{equation}
\begin{split}
\bar{J}\mathcal{G}J & =\sum_{n}\int d^{3}k\begin{pmatrix}
\bar{J}(\omega_{n},\vec{k}) & J(-\omega_{n},-\vec{k})
\end{pmatrix}\\
 & \times\frac{4}{\omega_{n}^{2}+\xi_{\vec{k}}^{2}(\chi_{\vec{k}}^{2}-4\eta_{\vec{k}}^{2})}\\
\times & \begin{bmatrix}
\frac{1}{2}(-i\omega_{n}-\xi_{\vec{k}}\chi_{\vec{k}}) & -\xi_{\vec{k}}\eta_{\vec{k}}\\
-\xi_{\vec{k}}\eta_{\vec{k}} & \frac{1}{2}(i\omega_{n}-\xi_{\vec{k}}\chi_{\vec{k}})
\end{bmatrix}\begin{pmatrix}
J(\omega_{n},\vec{k})\\
\bar{J}(-\omega_{n},-\vec{k}) 
\end{pmatrix}
\end{split}
\end{equation}
Note that for $\theta_{\vec{k}}=0, \ \forall\vec{k}$ which is equivalent to saying we have not done Bogoliubov transformation in that case we can get back the known result that the 2-point correlation function is given by single quantity $<\bar{\phi}(\omega_{n},\vec{k})\phi(\omega_{n},\vec{k})>\propto\frac{1}{i\omega_{n}-\xi_{\vec{k}}}$.\\[5pt]
Now let's do the matrix multiplication and write down the $\bar{J}\mathcal{G}J$ explicitly 
\begin{equation}
\begin{split}
\bar{J}\mathcal{G}J & =\sum_{n}\int d^{3}k\frac{4}{\omega_{n}^{2}+\xi_{\vec{k}}^{2}}\Big[\bar{J}(\omega_{n},\vec{k})(-i\omega_{n}-\xi_{\vec{k}}\chi_{\vec{k}})J(\omega_{n},\vec{k})\\
 & -\xi_{\vec{k}}\eta_{\vec{k}}(\bar{J}(\omega_{n},\vec{k})\bar{J}(-\omega_{n},-\vec{k})+J(\omega_{n},\vec{k})J(-\omega_{n},-\vec{k}))\Big]
\end{split}
\end{equation}
Note that for this case we found out that non-vanishing 2-point functions are
\begin{equation}
\begin{split}
<\phi(\omega_{n},\vec{k})\phi(-\omega_{n},-\vec{k})> & =-\frac{8\xi_{\vec{k}}\eta_{\vec{k}}}{\omega_{n}^{2}+\xi_{\vec{k}}^{2}}\\
<\bar{\phi}(\omega_{n},\vec{k})\bar{\phi}(-\omega_{n},-\vec{k})> & =-\frac{8\xi_{\vec{k}}\eta_{\vec{k}}}{\omega_{n}^{2}+\xi_{\vec{k}}^{2}}\\
<\bar{\phi}(\omega_{n},\vec{k})\phi(\omega_{n},\vec{k})> & =\frac{4(-i\omega_{n}-\xi_{\vec{k}}\chi_{\vec{k}})}{\omega_{n}^{2}+\xi_{\vec{k}}^{2}}
\end{split}
\end{equation}
Note that our result not only matches with known result in appropriate limit but also consistent with the fact that the first two result should be complex conjugate to each other, and here it's trivially satisfied becuase of the fact that we consider $\eta_{\vec{k}},\chi_{\vec{k}}$ are real functions of $\vec{k}$ which we have assumed from the begining of the calculation for sake of convenience. Note also that non-zero value of the first two correlation function is a consequence of violation of particle number conservation.\\[5pt]
\subsection{Evolution of number of particle in a state}
To get to know that number of particles containing in states in new Fock space in free-field theory is not conserved one can also check following quantity which one can get from the Hamiltonian defined in new Fock space. Note that according to eq.(25)
\begin{equation}
\begin{split}
\frac{d}{dt}(\hat{c}_{\vec{k}}^{\dagger}\hat{c}_{\vec{k}}) & =i[\hat{H},\hat{c}_{\vec{k}}^{\dagger}]\hat{c}_{\vec{k}}+i\hat{c}_{\vec{k}}^{\dagger}[\hat{H},\hat{c}_{\vec{k}}]\\
 & =i\int d^{3}l \ \varepsilon_{\vec{l}}\Big[[(\chi_{\vec{l}}\hat{c}_{\vec{l}}^{\dagger}\hat{c}_{\vec{l}}+\eta_{\vec{l}}(\hat{c}_{-\vec{l}}\hat{c}_{\vec{l}}+\hat{c}_{\vec{l}}^{\dagger}\hat{c}_{-\vec{l}}^{\dagger})),\hat{c}_{\vec{k}}^{\dagger}]\hat{c}_{\vec{k}}\\
 & +\hat{c}_{\vec{k}}^{\dagger}[(\chi_{\vec{l}}\hat{c}_{\vec{l}}^{\dagger}\hat{c}_{\vec{l}}+\eta_{\vec{l}}(\hat{c}_{-\vec{l}}\hat{c}_{\vec{l}}+\hat{c}_{\vec{l}}^{\dagger}\hat{c}_{-\vec{l}}^{\dagger})),\hat{c}_{\vec{k}}]\Big]\\
 & =i\varepsilon_{\vec{k}}\Big[\chi_{\vec{k}}\hat{c}_{\vec{k}}^{\dagger}\hat{c}_{\vec{k}}+2\eta_{\vec{k}}\hat{c}_{-\vec{k}}\hat{c}_{\vec{k}}-\chi_{\vec{k}}\hat{c}_{\vec{k}}^{\dagger}\hat{c}_{\vec{k}}-2\eta_{\vec{k}}\hat{c}_{\vec{k}}^{\dagger}\hat{c}_{-\vec{k}}^{\dagger}\Big]\\
 & =2i\eta_{\vec{k}}(\hat{c}_{-\vec{k}}\hat{c}_{\vec{k}}-\hat{c}_{\vec{k}}^{\dagger}\hat{c}_{-\vec{k}}^{\dagger})\neq0
\end{split}
\end{equation}
Now if we consider a state say $\ket{\psi}$ in new Fock space then 
\begin{equation}
\begin{split}
\frac{d}{dt}\bra{\psi} & \hat{c}_{\vec{k}}^{\dagger}\hat{c}_{\vec{k}}\ket{\psi}\\
 & =i\bra{\psi}[\hat{H},\hat{c}_{\vec{k}}^{\dagger}]\hat{c}_{\vec{k}}+i\hat{c}_{\vec{k}}^{\dagger}[\hat{H},\hat{c}_{\vec{k}}]\ket{\psi}\\
 & =2i\eta_{\vec{k}}\bra{\psi}(\hat{c}_{-\vec{k}}\hat{c}_{\vec{k}}-\hat{c}_{\vec{k}}^{\dagger}\hat{c}_{-\vec{k}}^{\dagger})\ket{\psi}\neq 0, \ \text{in general}\\
 & =2i\eta_{\vec{k}}[\bra{\psi}\hat{c}_{-\vec{k}}\hat{c}_{\vec{k}}\ket{\psi}-\bra{\psi}\hat{c}_{-\vec{k}}\hat{c}_{\vec{k}}\ket{\psi}^{*}]\\
 & =-4\eta_{\vec{k}}\text{Im}[\bra{\psi}\hat{c}_{-\vec{k}}\hat{c}_{\vec{k}}\ket{\psi}]\\
\implies\frac{d}{dt}\bra{\psi} & \sum_{\vec{k}}\hat{c}_{\vec{k}}^{\dagger}\hat{c}_{\vec{k}}\ket{\psi}=\frac{d}{dt}\bra{\psi}\hat{N}\ket{\psi}\neq0 
\end{split}
\end{equation}
which will be clear if one use completeness relation of occupation number basis of new Fock space. This clearly shows violation of particle number conservation.\\[5pt]
Recall I have mentioned earlier that number operator is a generator of phase in coherent states, in the above calculation we can alearly see that number operator does not commute with the Hamiltonian under Bogoliubov transformation $[\hat{H},\hat{N}]\neq0$, therefore phase generator or the charge corresponfing to U(1) symmetry is broken. Because of such non-commutativity Hamiltonian and number operator don't have simultaneous eigenstates. For example we can clearly see that vacuum state which is defined to be a state that is annihilated by annihilation operator is no longer an eigenstate of Hamiltonian operator although it conatains zero number of particles since vacuum expectation value of Hamiltonian is zero. And we can also check that there is no particle production happen in vacuum state since
\begin{equation}
\begin{split}
\frac{d}{dt}\bra{0(\theta)} & \hat{c}_{\vec{k}}^{\dagger}\hat{c}_{\vec{k}}\ket{0(\theta)}\\
=2i\eta_{\vec{k}}\bra{0(\theta)} & (\hat{c}_{-\vec{k}}\hat{c}_{\vec{k}}-\hat{c}_{\vec{k}}^{\dagger}\hat{c}_{-\vec{k}}^{\dagger})\ket{0(\theta)}=0\\
\implies\frac{d}{dt}\bra{0(\theta)} & \hat{N}\ket{0(\theta)}=0
\end{split}
\end{equation}
Note that saying particle creation out of vacuum by showing that old vacuum state in old Fock space representation is not true vacuum in new Fock space(which is mostly done \cite{degner2010cosmological}, \cite{Ford:1997hb}, \cite{biswas1995particle}, \cite{book:274311}, \cite{hossenfelder2003particle}, \cite{0305-4470-8-4-022}, \cite{ford1987gravitational}, \cite{birrell1980massive}, \cite{winitzki2005cosmological}, \cite{frieman1989particle}) is not equivalent of saying violation of particle number conservation because old vacuum state and multi-particle states in old Fock space do not belong to new Fock space in field theory because of infinite volume limit(even is some cases people consider amplitude of particle propagation under time evolution in path integral formalism by considering a initial and final states to be vacuum states at different time \cite{chitre1977path}, \cite{duru1986particle} which is also not a correct description). Therefore, one should carefully choose action of operactors according to their domain.\\[5pt]
Now let's check whether or not the vacuum state remains vacuum states under infinitesimal time evolution
\begin{equation}
\begin{split}
\hat{c}_{\vec{k}}e^{-i\hat{H}\epsilon}\ket{0(\theta)} & =e^{-i\hat{H}\epsilon}e^{i\hat{H}t}\hat{c}_{\vec{k}}e^{-i\hat{H}\epsilon}\ket{0(\theta)}\\
=e^{-i\hat{H}\epsilon} & \Big[\hat{c}_{\vec{k}}+i\epsilon[\hat{H},\hat{c}_{\vec{k}}]+\mathcal{O}(\epsilon^{2})\Big]\ket{0(\theta)}\\
=e^{-i\hat{H}t} & \Big[i\epsilon[\hat{H},\hat{c}_{\vec{k}}]+\mathcal{O}(\epsilon^{2})\Big]\ket{0(\theta)}\\
=e^{-i\hat{H}t} & [(-\chi_{\vec{k}}\hat{c}_{\vec{k}}-2\eta_{\vec{k}}\hat{c}_{-\vec{k}}^{\dagger})+\mathcal{O}(\epsilon^{2})]\ket{0(\theta)}\neq\ket{0(\theta)}
\end{split}
\end{equation}
So, we can see that even under infinitesimal time evolution vacuum state is no longer vacuum state of new transformed Fock space.\\[5pt]
Note also that in this whole setup we have not considered about action functional of this theory. From the action one can easily notice that since it is a real scalar field theory there is no breaking of $U(1)$-symmetry at all but once we write down the Hamiltonian and define what would be proper action of U(1) to check not charge conservation but rather particle number conservation because charge conservation may not violate because from vacuum state one can produce particle-antiparticle pairs but since for a real scalar field charge is zero we don't have to worry about charge conservation.

\subsection{Non-invariance under U(1) action in an example of 2-particle scattering interaction}
Now consider that system is interacting with the following interaction term in old Fock space represenatation
\begin{equation}
\hat{H}_{\text{int}}=\sum_{\vec{q},\vec{k},\vec{l}}v(\vec{q})\hat{a}_{\vec{k}+\vec{q}}^{\dagger}\hat{a}_{\vec{l}-\vec{q}}^{\dagger}\hat{a}_{\vec{l}}\hat{a}_{\vec{k}}
\end{equation}
where $v(\vec{q})$ is the interaction strength which depends on the exchanged momentum 2 particle scattering process at tree-level.\\[5pt]
If we do Bogolibuov transformation(then taking thermodynamic limit since we are considering field theory) which is equivalent to switch on gravity then in new Fock space representation we will get following interaction term
\begin{equation}
\begin{split}
\hat{H}_{\text{int}}=\sum_{\vec{q},\vec{k},\vec{l}}v(\vec{q}) & \Big[(\cosh\theta_{\vec{k}+\vec{q}}\hat{c}_{\vec{k}+\vec{q}}^{\dagger}+\sinh\theta_{\vec{k}+\vec{q}}\hat{c}_{-\vec{k}-\vec{q}})\\
 & \times(\cosh\theta_{\vec{l}-\vec{q}}\hat{c}_{\vec{l}-\vec{q}}^{\dagger}+\sinh\theta_{\vec{l}-\vec{q}}\hat{c}_{-\vec{l}+\vec{q}})\\
\times(\cosh\theta_{\vec{l}}\hat{c}_{\vec{l}}+\sinh\theta_{\vec{l}} & \hat{c}_{-\vec{l}}^{\dagger})\times(\cosh\theta_{\vec{k}}\hat{c}_{\vec{k}}+\sinh\theta_{\vec{k}}\hat{c}_{-\vec{k}}^{\dagger})\Big]\\
=\cosh\theta_{\vec{k}+\vec{q}}\cosh\theta_{\vec{l}-\vec{q}} & \cosh\theta_{\vec{l}}\cosh\theta_{\vec{k}}\hat{c}_{\vec{k}+\vec{q}}^{\dagger}\hat{c}_{\vec{l}-\vec{q}}^{\dagger}\hat{c}_{\vec{l}}\hat{c}_{\vec{k}}\\
+\cosh\theta_{\vec{k}+\vec{q}}\cosh\theta_{\vec{l}-\vec{q}} & \cosh\theta_{\vec{l}}\sinh\theta_{\vec{k}}\hat{c}_{\vec{k}+\vec{q}}^{\dagger}\hat{c}_{\vec{l}-\vec{q}}^{\dagger}\hat{c}_{\vec{l}}\hat{c}_{-\vec{k}}^{\dagger}\\
+\cosh\theta_{\vec{k}+\vec{q}}\cosh\theta_{\vec{l}-\vec{q}} & \sinh\theta_{\vec{l}}\cosh\theta_{\vec{k}}\hat{c}_{\vec{k}+\vec{q}}^{\dagger}\hat{c}_{\vec{l}-\vec{q}}^{\dagger}\hat{c}_{-\vec{l}}^{\dagger}\hat{c}_{\vec{k}}\\
+\cosh\theta_{\vec{k}+\vec{q}}\cosh\theta_{\vec{l}-\vec{q}} & \sinh\theta_{\vec{l}}\sinh\theta_{\vec{k}}\hat{c}_{\vec{k}+\vec{q}}^{\dagger}\hat{c}_{\vec{l}-\vec{q}}^{\dagger}\hat{c}_{-\vec{l}}^{\dagger}\hat{c}_{-\vec{k}}^{\dagger}\\
+\cosh\theta_{\vec{k}+\vec{q}}\sinh\theta_{\vec{l}-\vec{q}} & \cosh\theta_{\vec{l}}\cosh\theta_{\vec{k}}\hat{c}_{\vec{k}+\vec{q}}^{\dagger}\hat{c}_{-\vec{l}+\vec{q}}\hat{c}_{\vec{l}}\hat{c}_{\vec{k}}\\
+\cosh\theta_{\vec{k}+\vec{q}}\sinh\theta_{\vec{l}-\vec{q}} & \cosh\theta_{\vec{l}}\sinh\theta_{\vec{k}}\hat{c}_{\vec{k}+\vec{q}}^{\dagger}\hat{c}_{-\vec{l}+\vec{q}}\hat{c}_{\vec{l}}\hat{c}_{-\vec{k}}^{\dagger}\\
+\cosh\theta_{\vec{k}+\vec{q}}\sinh\theta_{\vec{l}-\vec{q}} & \sinh\theta_{\vec{l}}\cosh\theta_{\vec{k}}\hat{c}_{\vec{k}+\vec{q}}^{\dagger}\hat{c}_{-\vec{l}+\vec{q}}\hat{c}_{-\vec{l}}^{\dagger}\hat{c}_{\vec{k}}\\
+\cosh\theta_{\vec{k}+\vec{q}}\sinh\theta_{\vec{l}-\vec{q}} & \sinh\theta_{\vec{l}}\sinh\theta_{\vec{k}}\hat{c}_{\vec{k}+\vec{q}}^{\dagger}\hat{c}_{-\vec{l}+\vec{q}}\hat{c}_{-\vec{l}}^{\dagger}\hat{c}_{-\vec{k}}^{\dagger}\\
+\sinh\theta_{\vec{k}+\vec{q}}\cosh\theta_{\vec{l}-\vec{q}} & \cosh\theta_{\vec{l}}\cosh\theta_{\vec{k}}\hat{c}_{-\vec{k}-\vec{q}}\hat{c}_{\vec{l}-\vec{q}}^{\dagger}\hat{c}_{\vec{l}}\hat{c}_{\vec{k}}\\
+\sinh\theta_{\vec{k}+\vec{q}}\cosh\theta_{\vec{l}-\vec{q}} & \cosh\theta_{\vec{l}}\sinh\theta_{\vec{k}}\hat{c}_{-\vec{k}-\vec{q}}\hat{c}_{\vec{l}-\vec{q}}^{\dagger}\hat{c}_{\vec{l}}\hat{c}_{-\vec{k}}^{\dagger}\\
+\sinh\theta_{\vec{k}+\vec{q}}\cosh\theta_{\vec{l}-\vec{q}} & \sinh\theta_{\vec{l}}\cosh\theta_{\vec{k}}\hat{c}_{-\vec{k}-\vec{q}}\hat{c}_{\vec{l}-\vec{q}}^{\dagger}\hat{c}_{-\vec{l}}^{\dagger}\hat{c}_{\vec{k}}\\
+\cosh\theta_{-\vec{k}-\vec{q}}\cosh\theta_{\vec{l}-\vec{q}} & \sinh\theta_{\vec{l}}\sinh\theta_{\vec{k}}\hat{c}_{-\vec{k}-\vec{q}}\hat{c}_{\vec{l}-\vec{q}}^{\dagger}\hat{c}_{-\vec{l}}^{\dagger}\hat{c}_{-\vec{k}}^{\dagger}\\
+\sinh\theta_{\vec{k}+\vec{q}}\sinh\theta_{\vec{l}-\vec{q}} & \cosh\theta_{\vec{l}}\cosh\theta_{\vec{k}}\hat{c}_{-\vec{k}-\vec{q}}\hat{c}_{-\vec{l}+\vec{q}}\hat{c}_{\vec{l}}\hat{c}_{\vec{k}}\\
+\sinh\theta_{\vec{k}+\vec{q}}\sinh\theta_{\vec{l}-\vec{q}} & \cosh\theta_{\vec{l}}\sinh\theta_{\vec{k}}\hat{c}_{-\vec{k}-\vec{q}}\hat{c}_{-\vec{l}+\vec{q}}\hat{c}_{\vec{l}}\hat{c}_{-\vec{k}}^{\dagger}\\
\end{split}
\end{equation}
\begin{align*}
\begin{split}
+\sinh\theta_{\vec{k}+\vec{q}}\sinh\theta_{\vec{l}-\vec{q}} & \sinh\theta_{\vec{l}}\cosh\theta_{\vec{k}}\hat{c}_{-\vec{k}-\vec{q}}\hat{c}_{-\vec{l}+\vec{q}}\hat{c}_{-\vec{l}}^{\dagger}\hat{c}_{\vec{k}}\\
+\sinh\theta_{\vec{k}+\vec{q}}\sinh\theta_{\vec{l}-\vec{q}} & \sinh\theta_{\vec{l}}\sinh\theta_{\vec{k}}\hat{c}_{-\vec{k}-\vec{q}}\hat{c}_{-\vec{l}+\vec{q}}\hat{c}_{-\vec{l}}^{\dagger}\hat{c}_{-\vec{k}}^{\dagger}
\end{split}
\end{align*}
Note that even the interaction term gets modified in curved spacetime such way that it violates particle number conservation because of the fact interaction is not invariant under the action of global $U(1)$ group but momentum conservation still holds. Remember we have chosen a curved spacetime where notion momentum modes are well-defined which means that spacetime line-element or metric is invariant under spatial translations.\\[5pt]
According to the non-vanishing 2-point function that we have found all the interaction terms in the interaction Hamiltonian in new Fock space representation contribute in 4-point and higher order correlation function in interacting theory which is easy to see if one follows perturbative approach.\\[5pt]
\subsection{Conclusion}
In the beginning of this article I emphasized on the fact that in thermodynamic limit although we have 2 disjoint vector spaces but still the we can do the canonical transformation. And we also restrict ourself to new Fock space because after taking infinite volume limit we can't get back to the old Fock space. In this article I am able to show how to look at the particle production phenomena under change in coordinate transformation or under frame change which is equivalent of doing Bogoliubov transformation in field theory in thermodynamic limit. We have shown how does change in frame breaks both global and local U(1) invariance which is suitably defined. We have also seen that there is no particle production happen out of vacuum state in new transformed Fock space under time evolution but it can happen out of other many-particle states and vacuum state is not an eigenvector of Hamltonian operator in transformed Fock space and vacuum state does not remain vacuum state under time evolution.
\section{Acknowledgement}
Author wants to thank Dr. Golam Mortuza Hossain, Gopal Sardar for helpful discussion regarding the subject matter and their comments on the idea of this paper. Author would also like to thank CSIR to support this work through JRF fellowship.

\bibliographystyle{plain}
\bibliography{bibtexfile}

\end{document}